\documentstyle[epsfig]{mn}
\def\lessapprox{\,\raise 0.6ex\hbox{$<$}\kern -0.75em\lower 0.47ex
    \hbox{$\sim$}\,}
\def\largapprox{\,\raise 0.6ex\hbox{$>$}\kern -0.75em\lower 0.47ex
    \hbox{$\sim$}\,}

\def\wth{\omega(\theta)}

\def\ol{\Omega_\Lambda}
\def\om{\Omega_m}
\def\be{\begin{equation}}
\def\ee{\end{equation}}
\def\hth{\hat{\phi}}
\begin{document}
\title[Angular Cross-Correlation of Galaxies]
{Angular Cross-Correlation of Galaxies: 
A Probe of Gravitational Lensing by Large-Scale Structure}
\author[R. Moessner and B. Jain]
{R. Moessner$^1$ and Bhuvnesh Jain$^{1,2}$\\
$^1$Max Planck Institut f\"{u}r Astrophysik, 
Karl Schwarzschild-Str. 1, 
85740 Garching, Germany\\
$^2$Department of Physics and Astronomy,
Johns Hopkins University, 
Baltimore, MD 21218, USA\\
}
\maketitle
\begin{abstract}
\noindent

The angular cross-correlation between two galaxy samples
separated in redshift is shown to be a useful measure of weak 
lensing by large-scale structure. Angular correlations in faint 
galaxies arise due to spatial clustering of the galaxies as 
well as gravitational lensing by dark matter along the 
line-of-sight. The lensing contribution to the 2-point 
auto-correlation function is typically small compared to 
the gravitational clustering. However the cross-correlation between 
two galaxy samples is nearly unaffected by gravitational clustering 
provided their redshift distributions do not overlap. The 
cross-correlation is then induced by magnification bias due to lensing 
by large-scale structure. We compute the expected amplitude of the 
cross-correlation for popular theoretical models of structure formation. 
For two populations with mean redshifts of $\simeq 0.3$ and $1$, we find 
a cross-correlation signal of $\simeq 1\%$ on arcminute scales and 
$\simeq 3\%$ on a few arcseconds. The dependence on the cosmological 
parameters $\Omega$ and $\Lambda$, on the dark matter power spectrum 
and on the bias factor of the foreground galaxy population is explored. 

\end{abstract}


\section{Introduction}

The angular auto-correlation function of galaxies has been
widely used to characterize the large-scale structure in the
universe (e.g. Peebles 1980). The observed 
galaxy distribution is well described by a
power law $\wth\propto \theta^{-\gamma}$, with slope 
$\gamma\simeq 0.8$. 

Gravitational lensing by large-scale structure along the line-of-sight
can alter the amplitude of $\wth$ (Gunn 1967). 
Lensing increases the area of a given patch on the sky, thus diluting
the number density. On the other hand, galaxies too faint to be 
included in a sample of given limiting magnitude are brightened due
to lensing and may therefore be included in the sample. The net 
effect, known as magnification bias, 
can go either way: it can lead to an enhancement or suppression of the 
observed number density of galaxies, depending on the slope of the
number-magnitude relation.  Variations in the number density which
are correlated over some angular separation alter $\wth$. 
Following Kaiser (1992) and Villumsen (1996), 
Moessner, Jain \& Villumsen (1997) (henceforth, MJV) have 
considered the effect of nonlinear gravitational evolution and 
magnification bias on $\wth$. MJV found
that for faint samples with mean redshift $z\simeq 1$ lensing
contributes $5-20\%$ of the signal, depending on the cosmological
model and angular scale. Since the lensing contribution is small
even for distant galaxies, it is difficult to interpret a measurement,
especially since it requires knowledge of the biasing of high-redshift
galaxies relative to the mass. 

In this paper we explore a different statistic, the cross-correlation
function of two different galaxy samples, in order to isolate the
effect of magnification bias. Consider two galaxy samples with
non-overlapping redshift distributions. If the minimal
distance between the two samples is several $100$ Mpc, the effects
of gravitational clustering are negligible. The cross-correlation
function in such a case is affected entirely by magnification
bias, and the dominant term is 
provided by the lensing effect of dark matter associated with
the foreground galaxy population. 
Cross-correlating observables affected by gravitational lensing,
such as image ellipticities of
galaxies at high redshift,
 with the positions of galaxies at lower redshift has 
proved to be fruitful for detecting the effect of gravitational
lensing, for example in galaxy-galaxy lensing. 
The cross-correlation of the number density of  high-redshift quasars 
with foreground galaxies has  also been  investigated 
observationally (Benitez \& Martinez-Gonzalez 1996) and 
theoretically (Bartelmann 1995, Dolag \& Bartelmann 1997).
The cross-correlation of magnification and shear has been
studied by Kaiser (1992), Sanz, Maritnez-Gonzales \& Benitez (1997) and
Schneider (1997), while the auto-correlation of the shear has been 
computed by  Blandford et al. (1991), Miralda-Escude (1991), Kaiser (1992), 
Bernardeau, van Waerbeke and Mellier (1996) and Jain \& Seljak (1997). 

Section 2 provides the formalism for computing the effects of
lensing and nonlinear gravitational evolution on the angular 
cross-correlation function of galaxies. Results for cold dark matter
(CDM)-like models are given in Section 3. We provide estimates of
the errors in observational estimates of the cross-correlation
in Section 4 and conclude in Section 5. 

\section{Angular cross--correlation: Formalism}

This section provides the formalism for computing the angular
correlation function of galaxies for a given cosmological model
and primoridal spectrum of fluctuations for the dark matter. 
The intrinsic gravitational clustering
as well as the effect of gravitational lensing
by the dark matter along the line-of-sight are included. 
Since we consider the angular correlation function, only the
projected quantities are observable. 

We will focus on the cross-correlation of two galaxy samples, one
at high redshift, $z\largapprox 1$, called the background sample, and the
other at low redshift, $z\lessapprox 0.5$, called the foreground sample. 
We include in our computations the true correlation between galaxies 
belonging to the high- and the low-redshift samples due to the overlap
of their redshift distributions. As the overlap in the two redshift 
distributions decreases, this intrinsic clustering decreases. 
The effect we wish to isolate is the apparent clustering induced by 
gravitational lensing via magnification bias.

Let $n_1(\hth)$ be the number density of galaxies belonging to
the  sample with a low mean redshift $\langle z_1\rangle$, observed in the 
direction  $\hth$ in the sky, and $n_2(\hth)$ that of the
sample with a higher mean redshift $\langle z_2\rangle > \langle z_1
\rangle $. The angular 
cross-correlation function is then
defined as 
\be
\wth=\left< \delta n_1(\hth) \delta n_2(\hth^\prime)   \right>  \; ,
\label{wcross}
\ee 
where
\be
\delta n_i(\hth) \equiv \frac{n_i(\hth)-{\bar{n}_i}}{{\bar{n}_i}}
\ee
and ${\bar{n_i}}$ is the average number density of the $i$th sample.
The fluctuation $\delta n_i$ arises due to the
true clustering of galaxies 
$\delta n_i^g(\hth)$, and due to magnification bias 
$\delta n_i^{\mu}(\hth)$,
\be
\delta n_i(\hth)=\frac{n_i(\hth)-{\bar{n}_i}}{{\bar{n}_i}}=\delta
n_i^g(\hth) +\delta n_i^{\mu}(\hth)  \; .
\label{ni}
\ee
The goal of this paper is to compute the cross-correlation $\wth$
arising from these two effects. 

For simplicity we assume a linear bias model where galaxies
trace the underlying dark matter fluctuations, 
%$\detla_g=\delta_{DM}$
\be
\delta_g(\vec{x}) = b \ \delta(\vec{x}) \; .
\ee
The fluctuations on the sky due to intrinsic clustering are a projection 
of the density fluctuations along the line-of-sight, weighted with
the bias factor and the radial distribution $W(\chi)$ of the galaxies 
\be
\delta n_i^g(\hth)=b_i \int_0^{\chi_H} d \chi W_i(\chi) \delta(r(\chi)
\hth, a) \; .
\ee
The metric, the comoving radial coordinate $\chi$ and the comoving 
angular diameter distance $r(\chi)$ are introduced 
in the Appendix. We assume for simplicity a constant bias factor 
independent of scale and redshift for each galaxy population.

To determine the fluctuation due to magnificaton bias consider the
logarithmic slope $s$ of the number counts of galaxies $N_0(m)$ in 
a sample with limiting magnitude $m$ (see MJV for details)
\be
s= \frac{d \log N_0(m)}{dm} \; .
\ee  
Magnification by amount $\mu$ changes the number counts to 
(e.g. Broadhurst, Taylor \& Peacock 1995)
\be 
N^\prime(m)=N_0(m) \mu^{2.5s-1} \, .
\label{mag1}
\ee
In the weak lensing limit the magnification is 
$\mu=1+2 \kappa$, where the convergence $\kappa$ is a weighted 
projection of the density field along the line-of-sight (see equation
\ref{kappa1} below). Since $\kappa\ll 1$ for weak lensing, 
equation \ref{mag1} for
the number counts reduces to
\be 
N^\prime(m)=N_0(m)\left[ 1+5(s-0.4)\kappa \right]\, .
\label{Np}
\ee
Using $g(\chi)$ to denote the radial weight function 
(e.g. Jain \& Seljak 1997) the convergence $\kappa$ is
\be
\kappa_i(\hth)=\frac{3}{2} \om \int_0^{\chi_H} d \chi g_i(\chi)
\frac{\delta(r(\chi) \hth,a)}{a} \; .
\label{kappa1}
\ee
The expression for $g(\chi)$ in terms of the $r(\chi)$ is given in 
the Appendix. 

Finally, using the above relations we can re-write equation \ref{ni}
for $\delta n_i(\hth)$ as
\be
\delta n_i(\hth)=\int_0^{\chi_H} d \chi\, f_i(\chi)\, \delta(r(\chi)
\hth, a) \; ,
\label{deln}
\ee
with
\be
f_i(\chi)= b_i \, W_i(\chi)+ 3 \, \om \, (2.5 s_i-1)\, \frac{g_i(\chi)}{a} \; .
\label{fi}
\ee
Inserting the above two equations into Eq.~\ref{wcross}, using the small-angle
approximation $\theta \ll 1$, and assuming  that the radial weight functions
$f_i(\chi)$ vary slowly compared to the scale of
density perturbations of interest gives (Villumsen 1996),
\begin{eqnarray}
\wth&=& 4 \pi^2 \int_0^{\chi_H} d \chi f_1(\chi) f_2(\chi) \nonumber \\
&&\times \int_0^\infty dk\, k\, P(\chi, k)\, J_0\left[k r(\chi)\theta\right]\, 
\delta(r(\chi)\hth, a) \; .
\end{eqnarray}
The power spectrum of dark matter fluctuations $P(\chi, k)$ is defined by 
\be
\left<\delta(\vec{k}) \delta^{*}(\vec{k^\prime}) \right>=(2
\pi)^6P(\chi, k) \delta(\vec{k}-\vec{k^\prime}) \; .
\ee

The angular cross-correlation function is composed of four terms.
In the case of $\langle z_2\rangle  > \langle z_1
\rangle $, these terms are 
\begin{eqnarray}
\wth&=&\left< \delta n_1^g(\hth) \delta n_2^g(\hth^\prime)   \right> +
     \left< \delta n_1^g(\hth) \delta n_2^{\mu}(\hth^\prime)   \right>
\nonumber \\
     &+& \left< \delta n_1^{\mu}(\hth) \delta n_2^{\mu}(\hth^\prime) \right>
      +\left< \delta n_1^{\mu}(\hth) \delta n_2^g(\hth^\prime)   \right>
\; . 
\label{4terms}
\end{eqnarray}
The first term is due to the intrinsic clustering of the galaxies
of the two samples  where their redshift distributions
overlap,
\begin{eqnarray}
\omega_{gg}(\theta)&=&
b_1 b_2 4 \pi^2 \int_0^{\chi_H} d \chi W_1(\chi) W_2(\chi)\nonumber \\
&&\times \int_0^\infty dk\, k\, P(\chi, k)\, J_0\left[k r(\chi)\theta\right] \;
.
\label{aug31}
\end{eqnarray}
Ideally we would like this term to be zero, in order to distinguish the
contribution due to lensing more clearly. This could be achieved 
by obtaining photometric redshifts for the galaxies, and selecting
two galaxy populations which do not overlap in their redshift
distributions. The cross-correlation of two such samples minimizes
the contribution due to intrisic clustering, which
removes uncertainties in the predictions due to  unknown
physical evolution of the galaxies.

The second term in equation \ref{4terms}
is due to the lensing of the background galaxies
by the dark matter in front of it, which is traced by the foreground
galaxies. The correlation thus induced between galaxies in the two
samples is given by,
\begin{eqnarray}
\omega_{gl}(\theta)&=&b_1 3 \om (2.5 s_2-1) 4 \pi^2 \int_0^{\chi_H} d \chi
W_1(\chi) \frac{g_2(\chi)}{a} \nonumber \\
&& \times \int_0^\infty dk\, k\, P(\chi, k)\,
J_0\left[k r(\chi)\theta\right] \; . 
\label{omegagl}
\end{eqnarray}

The third term is due to dark matter in front of both of the
galaxy samples doing the lensing. The fourth is due to dark matter traced
by the background galaxies lensing the foreground galaxies. It is
non-zero only if there
is an overlap in the redshift distributions of the two samples. 
For all cases of interest these two terms are negligible compared
to the second term, $\omega_{gl}$. 

\subsection{Dependence on the cosmological model}

Equation 15 shows how  $\wth$ depends linearly on $\om$, aside from
the dependences on $\om$ and $\ol$ contained in the line-of-sight integral. 
These arise from two sources: (i) the distance factors contained in
$J_0$, $g(\chi)$ and $W(\chi)$, and (ii) 
the growth and amplitude of the  power spectrum. 
In the linear regime, $P(\chi, k)$ depends on the linear 
growing mode of density perturbations $D(\chi)$ and the 
normalization $\sigma_8$ (which in turn can
depend on the cosmological parameters) as
\be
P(\chi,k) \sim \left[ \sigma_8 D(\chi)\right]^2 \; .
\ee
The linear growing mode is well approximated by 
(Carroll, Press \& Turner 1992) 
\begin{eqnarray}
D(\chi)&=&\frac{5}{2}\, a \, \Omega(a) \
[\Omega(a)^{4/7}- \lambda(a)  \nonumber \\
&&+ (1+\Omega(a)/2)(1+\lambda(a)/70) ]^{-1}  \; , 
\end{eqnarray}
where we have defined, following Mo, Jing \& B\"orner (1996), the time dependent fractions of density in matter and
vacuum energy, $\Omega(a)$ and $\lambda(a)$ in terms of the present-day
values $\om$ and $\ol$,
\be
\Omega(a)=\frac{\om}{a+\om(1-a)+\ol(a^3-a)}
\ee
and
\be
\lambda(a)=\frac{a^3\ol}{a+\om(1-a)+\ol(a^3-a)} \; .
\ee
Moreover, the spatial 
geometries differ in different models, leading to a dependence of
the angular distance $r(\chi)$ on $\om$ and $\ol$ according to Eq.~\ref{rchi}. 

The redshift distribution of galaxies can be modelled by
\begin{equation}
n(z)=\frac{\beta z^2}{z_0^3 \Gamma[3/\beta]} 
\exp\left[-(z/z_0)^{\beta}\right] \;
,
\label{nz}
\end{equation}
for $\beta=2.5$, which agrees reasonably well with the redshift
distribution estimated 
for the Hubble deep field from photometric redshifts (Mobasher et al. 1996). 
The mean redshift is then given by 
\begin{equation}
\langle z\rangle=\frac{\Gamma(4/\beta)}{\Gamma(3/\beta)}\ z_0  \; .
\end{equation}

The four different cosmological models we consider are
a flat universe with $\Omega_m=1$, an open
model with $\Omega_m=0.3$, and a flat $\Lambda-$dominated 
model, with $\Omega_m=0.3$ and
$\Omega_{\Lambda}=0.7$. These three models are  normalized to cluster 
abundances (White, Efstathiou \& Frenk 1993; Viana \& Liddle 1995; 
Ecke, Cole \& Frenk 1996; Pen 1996),
\begin{equation}
\sigma_8\, \simeq\,  0.6 \, \Omega_m^{-0.6} \; .
\end{equation}
(For $\Omega_m=0.3$, we use $\sigma_8=1$, which is close to the
results of the approximate formula given above.)
Our fourth model is  a flat universe, $\om=1$,  with a high normalization of 
$\sigma_8=1$.

The nonlinear power spectrum is obtained from
the linear one through the fitting formulae of Jain, Mo \& White (1995) for
$\om=1$, and from those of Peacock \& Dodds (1996) for the open and $\Lambda-$
dominated models.
These fitting formulae are based on the idea of relating the
nonlinear power spectrum at scale $k$ to the linear power spectrum at 
a larger scale $k_L$, where the relation $k(k_L)$ depends on
the power spectrum itself (Hamilton et al. 1991). 
They have been calibrated from and tested extensively against N-body 
simulations. For the linear power spectrum we take a CDM-like spectrum
with shape parameter $\Gamma=0.25$, which provides a good fit to
observations. The transfer function for the initially scale-invariant
power spectrum is taken from Bardeen et al. (1992).

\section{Cross-correlation function for CDM-like power spectra}
\label{wtotal}

\begin{figure*}
 \centerline{\psfig{figure=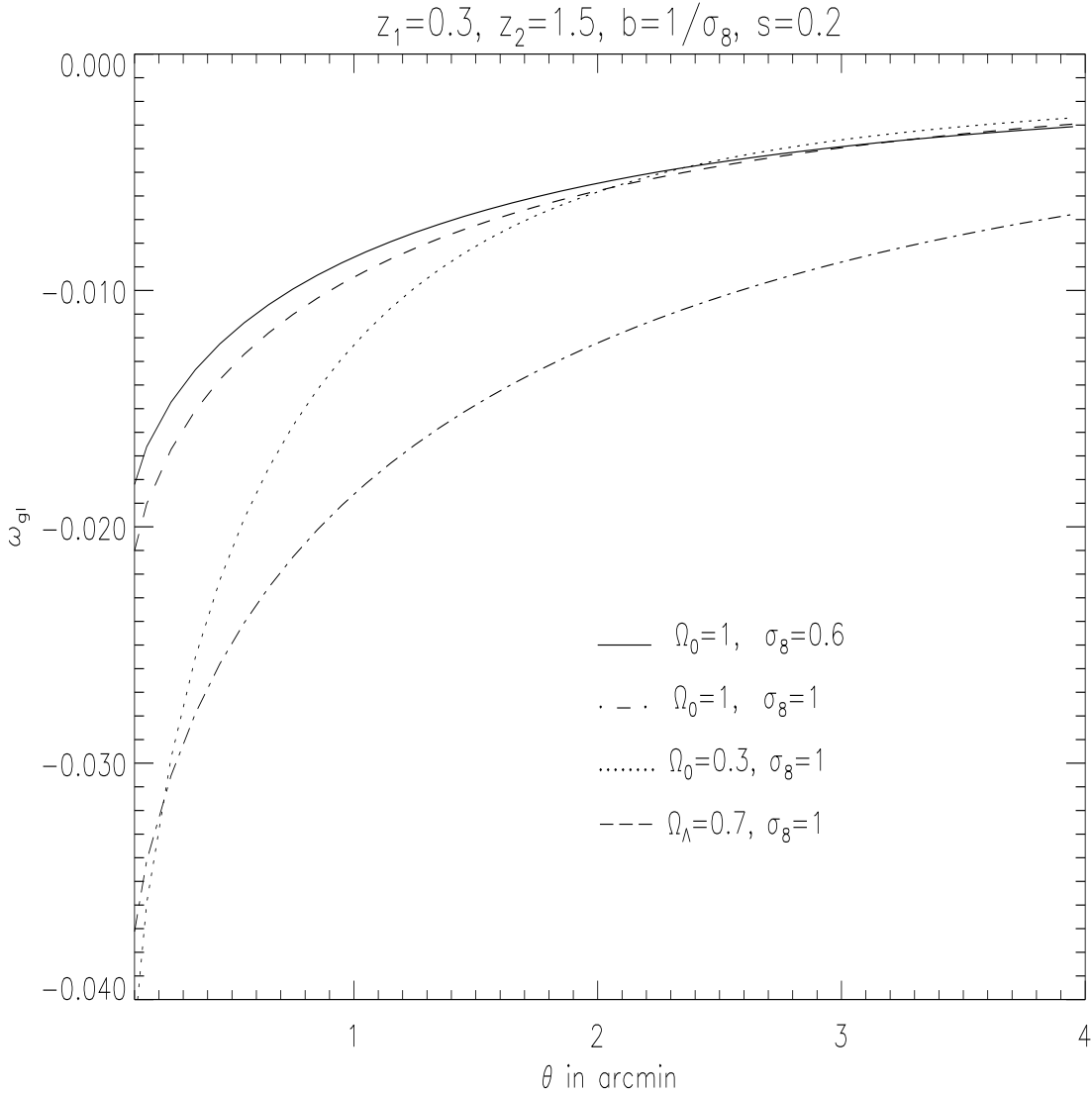,height=6in,width=6in}}
 \caption{The cross-correlation function $\omega_{gl}$ 
as a function of $\theta$
is shown for the four cosmological models. Since the number count
slope $s=0.2$, magnification bias induces an
anti-correlation between the foreground and background sample. Hence
the sign of $\omega_{gl}$ is negative. The differences between the
four models are discussed in the text.}
 \label{wth}
\end{figure*}

For samples not overlapping in their redshift distributions, the 
cross--correlation $\omega_{gl}$ induced by lensing 
dominates the contribution to the cross-correlation function. 
The signal due to intrinsic clustering is negligibly small,
though if the redshift distributions overlap significantly, it
would swamp the lensing signal. 
In this section we present results on the cross-correlation of
two galaxy samples with non-overlapping redshift distributions. 
 
We computed $\omega_{gl}(\theta=1^\prime)$ for two populations with mean
redshifts of $0.3$ and $1.5$ using the redshift distribution $n(z)$ of
Eq.~\ref{nz}. We found that the results differ by
$5 - 10 \%$ from those obtained using a delta--function redshift distribution. 
Therefore we will use the simpler form of a 
delta--function distribution for computing  $\omega_{gl}(\theta)$.
For calculating contributions due to intrinsic clustering, however,
it is necessary to use the full $n(z)$ of Eq.~\ref{nz}. 
We do this in Section~\ref{err} below to estimate 
the relative error made 
by misidentifying  background galaxies as foreground ones. 

In Figure~\ref{wth} we plot $\omega_{gl}$ as a function of $\theta$
for  a  foreground galaxy sample at $z_1=0.3$ and a background sample at
$z_2=1.5$, for the four cosmological models described in section 2.1.
We choose a bias factor of $b=1/\sigma_8$, which is 
in agreement with large--scale galaxy clustering data; $\omega_{gl}$
is proportional to $b$. 
We assume a number count slope of $s=0.2$ for 
the background sample. 
For $s<0.4$ the induced correlations are negative since 
$\omega_{gl} \omega_{gl}\propto (s-0.4)$. This linear relation 
also makes it simple to scale our results to other values of $s$. 
A sample with a slope close to $0.2$ may be obtained by defining 
color selected subsamples, using the fact that the number count slope is a 
decreasing function of $V-I$ color (Villumsen, Freudling \& da Costa 1996; 
Broadhurst et al. 1997). The price of selecting a subsample is a smaller
number of galaxies and therefore larger Poisson errors; so in practice 
a careful cut suited to the available data would need to be made. 

Figure~\ref{wth} shows that the typical cross-correlation signal
expected on sub-arcminute scales is a few percent. On angular scales
larger than 1 arcminute the signal drops to less than a percent. 
For fixed $\sigma_8$ it is largest for the Einstein-de Sitter model 
and smallest for the cosmological constant model, at least on 
arcminute scales or smaller. However if $\sigma_8$ is determined
from cluster-abundances, the cross-correlation on small scales 
is largest for the open model. 

\begin{figure}
 \centerline{\psfig{figure=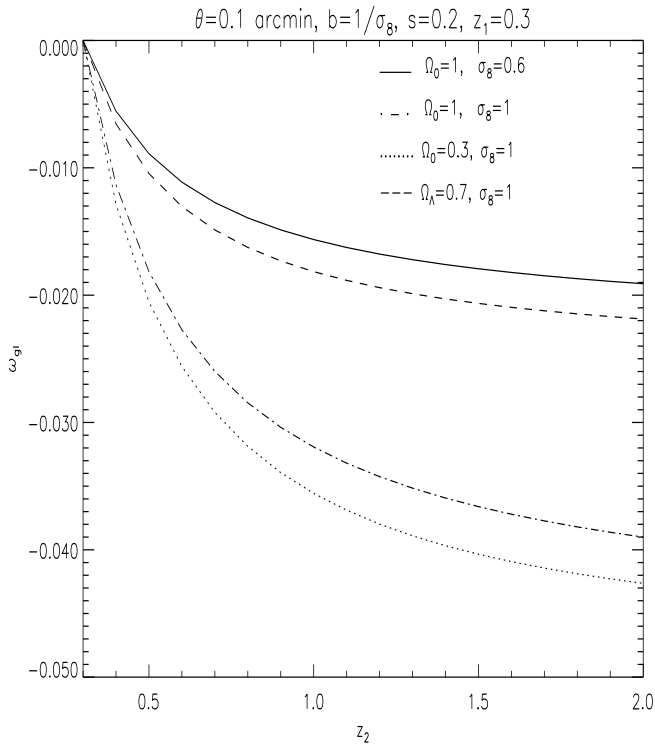,height=4in,width=3.5in}}
 \caption{Cross-correlation function as a function of
$z_2$ is shown for $z_1=0.3$ at $\theta=0.1'$ for the four cosmological models.
The negative correlations induced by magnification bias become
stronger with $z_2$, but do not change much beyond a redshift of 1. }
 \label{wth01}
\end{figure}

\begin{figure}
 \centerline{\psfig{figure=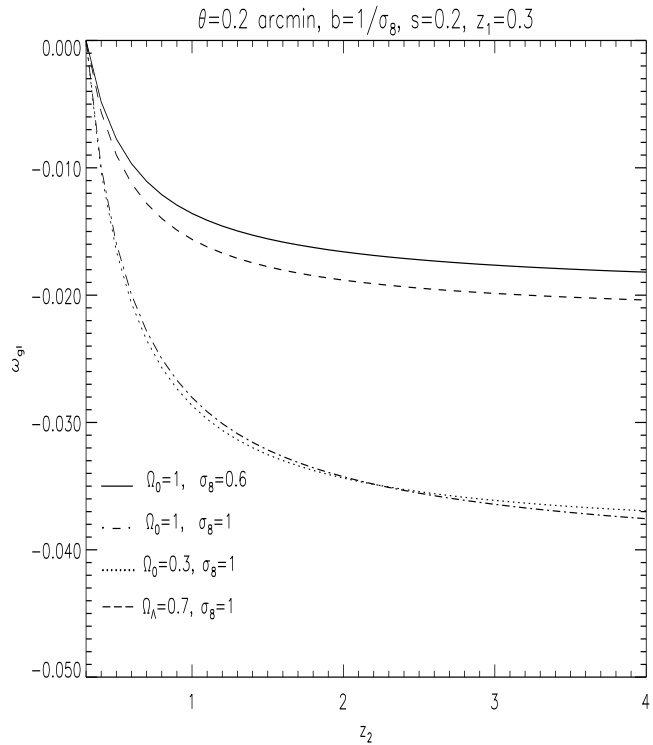,height=4in,width=3.5in}}
 \caption{Cross-correlation function as a function of
$z_2$ is shown for $\theta=0.2'$ for the four cosmological models.
All parameters except $\theta$ are as in the previous figure. }
 \label{wth02}
\end{figure}

The dependence of the cross-correlation
on the redshift of the background sample is shown in Figure~\ref{wth01}. 
We have plotted the cross--correlation function 
$\omega_{gl}(\theta=0.1^\prime)$
for a  foreground galaxy sample
at $z_1=0.3$ as a function of the redshift $z_2$ of the  background
sample. Figure~\ref{wth02} is the same as Figure~\ref{wth01}, but for 
$\theta=0.2^\prime$. These figures show the slow increase in the 
amplitude of the
signal with $\langle z_2\rangle$ above a redshift of 1. There is 
no significant variation in the shape of the curves among the four 
cosmological models. Thus if the amplitudes were normalized to the same 
value at $z_2\simeq 1$, there would be very little difference between 
the curves at higher $z_2$. 

The difference between the predictions of the four models shown
in Figures~\ref{wth}-\ref{wth02} can be qualitatively understood 
as follows. The dominant dependence arises due to the factor of
$\om$ outside the integral in equation~\ref{omegagl} for $\omega_{gl}$. 
Taking the normalization of the power spectrum into account
this is reduced to $\om^{0.4}$ for the cluster-abundance normalization
of $\sigma_8$ and the bias relation $b=1/ \sigma_8$. This is because
the expression for $\omega_{gl}$ depends explicitly on the factor  
$b \om \sigma_8^2$. 

The line-of-sight integral in equation~\ref{omegagl} further
weakens the dependence on $\om$. 
In  a low-$\om$ universe, the growth of perturbations is slowed down at
late times. Hence, normalizing
to present-day cluster abundances leads to a higher
normalization at earlier times compared to the $\om=1$ models. This
in turn means that nonlinear effects, which are significant
on angular scales of  $1^\prime$ or less, 
become important earlier on
and lead to a larger enhancement due to nonlinear clustering by today.
The nonlinear enhancement is reinforced by a geometrical effect: 
for lower $\om$, and even more so for larger $\Omega_\Lambda$ 
(for given $\om$),  the physical distance to a
given redshift is larger. This leads to a larger lensing path-length 
and thus a further increase in the lensing signal. 

The combination of all these effects is shown in Figure~\ref{wth}. 
On scales well below an arcminute the signal for the open model 
becomes comparable to that in the Einstein-de Sitter model. 
This is due to the dominance of the nonlinear enhancement on 
these scales. The curve for the open model is also distinctly
steeper than for the others. For $\theta > 2^\prime$, however, there is
no significant variation in the shape of the curves. 
For the $\Lambda-$dominated
model the effect of the growing mode is not as strong as for the open
model. Thus on small scales, even though the geometric effect gives a stronger
enhancement than for the open model, the net amplitude of
$\omega_{gl}$ is smaller than for the open model.

Finally, note that $\omega_{gl}$ is proportional to the bias factor of the 
foreground sample. If this bias factor is larger or smaller than the 
value of $b=1/\sigma_8$ which we assumed to fit the large--scale
structure data, $\omega_{gl}$ will be correspondingly altered. 
 
\section{Estimate of errors}
\label{err}

\begin{figure*}
 \centerline{\psfig{figure=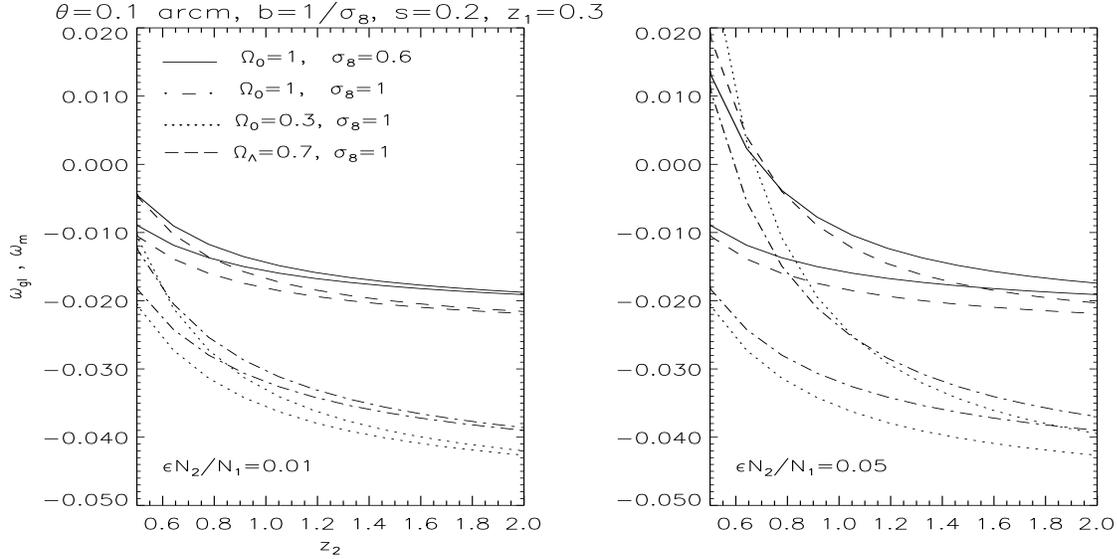,height=3in,width=6in}}
 \caption{True cross-correlation function $\omega_{gl}$ (lower set of 
matching curves) and 
measured cross-correlation function $\omega_{m}$ (upper set of
 matching curves) when 
$\epsilon \%$ of the background galaxies are misidentified as
foreground ones, as a function of
$z_2$, for $\theta=0.1'$, $z_1=0.3$ and the four cosmological models.}
 \label{wmiss}
\end{figure*}

We consider two sources of error involved in an observational
determination of the cross-correlation from two galaxy samples. 
The first potential error arises when a background galaxy is
mis-identified as a foreground galaxy. In this case the auto--correlation
of the background galaxy sample will erroneously enter into the
measured cross-correlation. Say $\epsilon \%$ of the background
galaxies' redshifts are sufficiently mis-estimated that they
are taken to be part of the foreground sample. 
The observed cross--correlation is then given by
\begin{equation}
 \omega_{m}= \omega_{gl} + \epsilon \frac{N_2}{N_1} \omega_{gg}(z_2) \, ,
\end{equation}
where the second term is the error made due to assigning galaxies to the 
wrong sample. It is proportional to the fraction of galaxies
which is mis-identified and to $\omega_{gg}(z_2)$, the
auto--correlation function evaluated
at the mean redshift of the background sample, for the redshift 
distribution of Eq.~\ref{nz}. We have calculated the auto--correlation 
at $1^\prime$ and 
scaled it to $0.1^\prime$ assuming a power law slope of $-0.8$;
$N_1$ and $N_2$ are the number of galaxies in the foreground and 
background samples.

In Figure~\ref{wmiss}  we plot the cross-correlation
$\omega_{gl}(\theta=0.1^\prime)$ (lower set of
 matching curves) together with the measured 
cross-correlation function $\omega_{m}$ (upper set of matching curves). 
Two values of $\epsilon$, $1\%$ and $5\%$ are used (for samples of 
equal size). The results show that
for $z_2>1$ the error is small for $\epsilon=1\%$ but not when
$\epsilon=5\%$. This sets an approximate standard required for using 
photometric redshifts or other possible methods to select the two 
galaxy samples. 

A second source of error is the statistical uncertainty in 
estimating the angular correlations. Using a Poisson distribution
to estimate the error in $\omega$ provides a rough
guide for the number of galaxies required to estimate the 
cross-correlation signal. 
The standard deviation $\delta \omega(\theta)$ 
in the estimate of $\omega(\theta)$ for a 
random distribution of objects is given by (Peebles 1980)
\begin{equation}
\delta \omega(\theta)^2 = 
\frac{1}{{N_1 N_2}} \, \frac{\Omega}{\delta \Omega}
\end{equation}
where $\Omega$ is the solid angle subtended by the survey area, and
$\delta\Omega$ is the fraction in the bin used for angle $\theta$. Note
that $\delta\omega^2$ is just the inverse of the number of
pairs in a given bin in $\theta$. 

For a sample with about $10^3$ galaxies per $0.01$ degree$^2$
(e.g. as in each field of Woods \& Fahlman (1997) 
whose sample reaches limiting magnitudes of 
$V\sim 25$, $R\sim 25$, $I\sim 24$), the above estimate gives 
$\delta\omega\simeq 4 \times 10^{-3}$ if 10 bins in $\theta$
are used. Thus in excess of about $10^3$ galaxies 
each in the foreground and background sample would be required for 
a detection of a $\largapprox 1 \%$
cross-correlation signal with a high level of significance. 

\section{Conclusions}

We have presented results for the cross-correlation of two galaxy samples
with different redshift distributions. 
The signal is dominated by the effect of magnification bias 
due to weak lensing. To ensure that the contribution from
gravitational clustering of the galaxies is negligible, we have assumed 
that the redshift distributions of the two samples do not overlap. 
With the use of photometric redshifts (e.g. Connolly et al. 1995; 
Sawicki, Lin \& Yee 1997), it is feasible to obtain
deep galaxy samples that can be separated into sub-samples
with desired redshift distributions. If only limiting magnitudes
are used to create two sub-samples, then there will be a significant
overlap and the interpretation of the signal is not as clean. 
Theoretical predictions can however be made for the expected signal
assuming a redshift distribution. 

The results shown in Figures 1-3 demonstrate that most models
predict a signal of 1-4\% for the cross-correlation function. 
These numbers apply for a background sample with a mean redshift
$\largapprox 1$ and a number count slope of $0.2$, on angular scales 
from a few arcseconds to an arcminute. As argued in section 4, the 
measurement of such a signal appears feasible in the near future. 

The cross-correlation function is a measure of the projected dark matter
power spectrum. For a given spectrum, the variation with angle on 
small scales is largest for open cosmological models and thus 
provides a probe of $\Omega$. The cross-correlation however is also
proportional to the bias factor of the foreground ($z\simeq 0.3-0.5$) 
galaxy sample. Thus there is a degeneracy in the dependance
on the cosmological model and the biasing of intermediate redshift 
galaxies. The bias factor can vary with the redshift of the
foreground sample and with $\Omega$. 
Given a model or empirical measure of the bias factor, 
the cross-correlation can be used as a probe
of the power spectrum and $\Omega$. Else, for a given cosmology, 
it can constrain the bias factor of galaxies. 
Ideally, by combining the cross-correlation  with 
other lensing measures which are independent of bias, 
such as the ellipticity auto--correlation function,
constraints on the cosmological model as well as on the
biasing of galaxies at intermediate redshifts can be obtained. 

\section*{Acknowledgement}

We are grateful to Simon White for many helpful suggestions. 
We would like to thank Matthias Bartelmann, Andrew Connolly, 
Peter Schneider, Alex Szalay and Jens Villumsen for stimulating 
discussions. The paper also benefitted from helpful comments 
by the referee, Andrew Taylor. 

\section*{Appendix A: Notation}

Following the notation of Jain \& Seljak (1997) we introduce
the unperturbed metric
\begin{equation}
ds^2=a^2(\tau)\left( -d\tau^2+ d\chi^2+r^2(d\theta^2+\sin^2 \theta d\phi^2)
 \right)  ,
\end{equation}
with $\tau$ being conformal time, and $\chi$ the radial 
comoving distance.  $\chi_H$ is used to denote 
the distance to the horizon. The
comoving angular diameter distance $r(\chi)$ is, 
\begin{eqnarray}
r(\chi)=\sin_K\chi \equiv
\left\{ \begin{array}{ll} K^{-1/2}\sin K^{1/2}\chi,\ K>0\\
\chi, \ K=0\\
(-K)^{-1/2}\sinh (-K)^{1/2}\chi,\ K<0 
\label{rchi}
\end{array}
\right.
\end{eqnarray}
where $K$ is the spatial curvature given by 
$K=-H_0^2(1-\Omega_m-\Omega-\Lambda)$ with $H_0$ being the 
Hubble parameter today. 

With $W(\chi)$ denoting the radial distribution 
of galaxies in the sample, the radial weight function $g(\chi)$ 
is given by
\begin{equation}
g(\chi)= r(\chi) \int_\chi^{\chi_H}
{r(\chi' -\chi) \over r(\chi')}W(\chi')d\chi'\ .
\end{equation}
For a delta-function distribution of galaxies,
$W(\chi')=\delta_D(\chi'-\chi_S)$, and $g(\chi)$ reduces to 
$g(\chi)=r(\chi)r(\chi_S-\chi)/r(\chi_S)$.

\end{document}